\documentclass[%
 reprint,
superscriptaddress,
 aps,
 prl
]{revtex4-2}

\usepackage{xfrac}
\usepackage{braket}
\usepackage{physics}
\usepackage{graphicx}
\usepackage{dcolumn}
\usepackage{bm}
\usepackage{subcaption}
\usepackage{graphicx}
\usepackage{amsmath}
\usepackage{amssymb}
\DeclareMathSizes{10}{10}{4}{4}
\usepackage[colorlinks=true, allcolors=blue]{hyperref}
\usepackage{floatrow}
\newcommand{\labelphantom}[1]{%
  \parbox{0pt}{\phantomsubcaption\label{#1}}%
}
\begin{document}
\preprint{APS/123-QED}

\title{Thermal quenching of classical and semiclassical scrambling}

\author{Vijay Ganesh Sadhasivam}\email{vgs23@cam.ac.uk}
\affiliation{Yusuf Hamied Department of Chemistry, University of Cambridge, Lensfield Road, Cambridge, CB2 1EW, United Kingdom}

\author{Andrew~C.~Hunt}
\thanks{A.C.H. and L.M. contributed equally to the article.}
\affiliation{Yusuf Hamied Department of Chemistry, University of Cambridge, Lensfield Road, Cambridge, CB2 1EW, United Kingdom}

\author{Lars Meuser}
\thanks{A.C.H. and L.M. contributed equally to the article.}
\affiliation{Yusuf Hamied Department of Chemistry, University of Cambridge, Lensfield Road, Cambridge, CB2 1EW, United Kingdom}
\affiliation{Department of Chemistry and Applied Biosciences, ETH Z\"urich (Swiss Federal Institute of Technology), Z\"urich 8093, Switzerland}

\author{Yair Litman}
\affiliation{Yusuf Hamied Department of Chemistry, University of Cambridge, Lensfield Road, Cambridge, CB2 1EW, United Kingdom}

\author{Stuart C.~Althorpe}
\affiliation{Yusuf Hamied Department of Chemistry, University of Cambridge, Lensfield Road, Cambridge, CB2 1EW, United Kingdom}

\def\figgy{1}

\date{\today}

\begin{abstract}
Quantum scrambling often gives rise to short-time exponential growth in out-of-time-ordered correlators (OTOCs). The scrambling rate over an isolated saddle point at finite temperature is shown here to be reduced by a hierarchy of quenching processes. Two of these appear in the classical limit, where escape from the neighbourhood of the saddle reduces the rate by a factor of two, and thermal fluctuations around the saddle reduce it further; a third process can be explained semiclassically as arising from quantum thermal fluctuations around the saddle, which are also responsible for imposing the Maldacena-Shenker-Stanford bound.
\end{abstract}

\maketitle

\section{Introduction}
Scrambling refers to the spreading of dynamical information as a result of local and global instabilities. In a quantum system, scrambling can be quantified using `out-of-time-ordered correlators' (OTOCs) \cite{larkin1969quasiclassical,maldacena2016bound,sekino2008fast} of the form
\begin{equation}\label{genOTOC}
    C(t) = \left \langle \left [\hat{W}(t), \hat{V}(0) \right]^{\dagger} \left [\hat{W}(t), \hat{V}(0) \right]  \right\rangle
\end{equation}
where $\hat{W}$ and $\hat{V}$
are hermitian operators representing local quantum information. For some systems, the OTOC grows exponentially at times $t$ shorter than the Ehrenfest time $\tau$, before flattening due to the onset of coherence. This short-time exponential growth is sometimes referred to as `quantum chaos' and the exponential growth rate
\begin{align}\lambda_Q = \frac{d \ln C(t)}{ dt}\end{align}
as a `quantum Lyapunov exponent'; this language is very loose, since exponential growth can occur in non-chaotic systems such as the barrier-scrambling systems we focus on below.
At sufficiently high temperatures, $\lambda_Q$ is close to the exponential growth rate $\lambda_\text{cl}$ of the corresponding classical phase-space average
\begin{align}\label{clOTOC}
    C_\text{cl}(t) = \hbar^2\left \langle \left\{W_t,V \right\}^2 \right \rangle
\end{align}
where $\{\cdot,\cdot\}$ denotes a Poisson bracket. For convenience we will often refer to expressions such as $C_\text{cl}(t)$  as `classical OTOCs'.
At lower temperatures, $\lambda_Q$ reduces in line with the Maldacena-Shenker-Stanford (MSS) bound \cite{maldacena2016bound}
\begin{align}\label{MSSbound}
    \lambda_Q(T) \leq \frac{2 \pi k_B T}{\hbar}
\end{align}
which has been shown to be of quantum statistical origin \cite{murthy2019bounds, tsuji,Pappalardi2022, sadhasivam2023instantons}.

{A significant body} of work on OTOCs has considered systems in which the scrambling is not classically chaotic, but is caused by unstable dynamics around an isolated saddle point. Examples of such systems include the Dicke \cite{DickeModelOTOC}, Bose-Hubbard  \cite{SPscramNotherm} and Lipkin-Meshkov-Glick (LMG) \cite{xu2020does} models, as well as simple barrier-crossing systems, in which the scrambling takes place between two wells separated by a barrier with imaginary frequency $\omega_b$ \cite{hashimoto2020exponential, bhattacharyya2021multi,sadhasivam2023instantons, zhang2022,
wolynes2023}. On the basis of \eqref{clOTOC}, one might expect that $\lambda_{\text{cl}}\simeq 2\omega_b$ and hence that $\lambda_Q$ also tends to this value at high temperatures; however, it is found \cite{hashimoto2020exponential, bhattacharyya2021multi,sadhasivam2023instantons, zhang2022,
wolynes2023} that $\lambda_{\text{cl}}\simeq \omega_b$. This factor-of-two reduction is passed on to $\lambda_Q$ and, together with instanton delocalisation over the barrier \cite{sadhasivam2023instantons}, is the reason that barrier-scrambling systems obey the MSS bound of  \eqref{MSSbound}.

In this article, we formalise the factor-of-two quenching of $\lambda_{\text{cl}}$, then investigate a further reduction of $\lambda_{\text{cl}}$ found to occur at short times. We show that this further reduction occurs because the exponential growth of the OTOC has not had sufficient time to become dominated by trajectories immediately at the saddle, but is instead an average over thermal fluctuations around the saddle. We then show that quantum thermal fluctuations analogously reduce $\lambda_{\text{Q}}$  over the short times  for which the quantum OTOC grows exponentially.

The factor-of-two reduction can be understood by considering the exponential growth of the classical OTOC at long times.
Similarly to ref.~\onlinecite{xu2020does}, we consider 
\begin{align}
    C^+_\text{cl}(t) = &\frac{\hbar^2}{hZ_{cl}} \int_{-\infty}^\infty\! da^+ \int_{-\infty}^\infty\! da^- \:e^{-\beta H(a^+,a^-)}  \left (\frac{\partial a^+_t}{\partial a^+} \right )^2_{\!a^-}\nonumber\\
\end{align}
 for a system with Hamiltonian
\begin{align}
    H(a^+,a^-) = \frac{a^+a^-}{2m} + G(a^+-a^-)
\end{align}
where
\begin{align}
    a^\pm = {1\over\sqrt{2}}\left(p\pm m\omega_b q\right) 
\end{align}
and $G(a^+-a^-)$ is the anharmonic part of the potential.  At long times, $C^+_\text{cl}(t)$ is dominated by the trajectories which have the fastest exponential growth, namely the trajectories which remain within a small region $|a^\pm|<\delta$ surrounding the saddle for all times up to $t$, for which ${\partial a^+_t}/{\partial a^+}=\exp(\omega t)$. We can therefore write  
\begin{align}
    C^+_\text{cl}(t) \underset{ t\to\infty\ }{\xrightarrow{\hspace{15 pt}}} &\frac{\hbar^2}{hZ_{cl}} \int_{-\infty}^\infty\! da^+ \int_{-\infty}^\infty\! da^- \:e^{-\beta H(a^+,a^-)} S(a^+)S(a^-)\nonumber\\
    &\quad\quad\quad\times S(a^+e^{\omega t})S(a^-e^{-\omega t})\left (\frac{\partial a^+_t}{\partial a^+} \right )^2_{\!a^-}\nonumber\\
    =&\frac{\hbar^2}{hZ_{cl}} \int_{-\delta e^{-\omega t}}^{\delta e^{-\omega t}}\! da^+ \int_{-\delta}^\delta\! da^- \:e^{-\beta H(a^+,a^-)} \left (\frac{\partial a^+_t}{\partial a^+} \right )^2_{\!a^-}
\end{align}
where $S(x) = 1$ for $-\delta < x < \delta$ and 0 otherwise, which implies that
\begin{align}\label{tinf_lamda}
\lim_{t\to\infty} \lambda_\text{cl} = \omega_b
\end{align}
(since one power of ${\partial a^+_t}/{\partial a^+}$ is cancelled by the exponentially shrinking limits on $a^+$).

For the short times $t<\tau$ over which the quantum OTOC grows exponentially, the behaviour of $C_\text{cl}(t)$ will be very far from the long time limit in \eqref{tinf_lamda}. Unless the barrier is sufficiently low, $C_\text{cl}(t)$ will not grow exponentially for $t<\tau$; and if it does grow, $C_\text{cl}(t)$ is likely to be dominated by a far wider spread of trajectories than those contained in $|a^\pm|<\delta$. Assuming that the effect of $G(a^+-a^-)$ is to reduce the negative Hessian away from the barrier \footnote{The function $G$ has been chosen to make the Hessian increase monotonically on either side of the barrier.}, we can expect that $\lambda_\text{cl}$ will be smaller than $\omega_b$ at short times, and that a corresponding reduction will affect the quantum OTOC (at least at temperatures $T>T_c$).

We investigated this behaviour numerically for the two-dimensional model potential
\begin{align}\label{Vcoupling}
    V(\textbf{q}) =  g\left ( x^2 -\frac{m\omega_b^2}{4g} \right)^2+ D(1-e^{-\alpha y})^2 + z^2\alpha^2x^2y^2 \slash 2
\end{align}
where $z$ is a tunable coupling parameter, $\textbf{q}=(x,y)$, $m=0.5$, $\omega_b=2$, $g=0.08$, $\alpha=0.382$, $D=3V_b$ and $V_b = m^2\omega_b^2\slash 16g$. This potential (shown in fig.~\ref{fig:fig2}) has a saddle-point at the origin and symmetric coupling between $x$ and $y$, the strength of which can be tuned by the parameter $z$ ($z=0$ corresponds to the 1D double well along $x$); the saddle imaginary frequency is $\omega_b$. 
We consider the quantum OTOC \footnote{Note that we use the Kubo regularisation as in eq.~17 of ref.~\onlinecite{sadhasivam2023instantons} for the quantum OTOC} as in \eqref{genOTOC} with $\hat{W} = \hat{x}$ and $\hat{V}=\hat{p}_x$,
\begin{align}\label{qqOTOC}
    C_\text{q}(t) = -\frac{1}{Z}\langle \left[\hat{x}(t), \hat{p}_x(0) \right]^2 \rangle
\end{align} 
for which the corresponding classical OTOC is
 \begin{align}
    C_\text{cl}(t) = \frac{1}{4\pi^2Z_{cl}} \int d\textbf{q} \: d\textbf{p} \: e^{-\beta H(\textbf{q},\textbf{p})} \left(\frac{\partial x_t}{\partial x}\right)^2_{\!(y,{\bf p})}
\end{align}

\if\figgy1
\begin{figure}[H]
    \labelphantom{fig1a}
    \labelphantom{fig1b}
    \centering
    \includegraphics[scale=0.7]{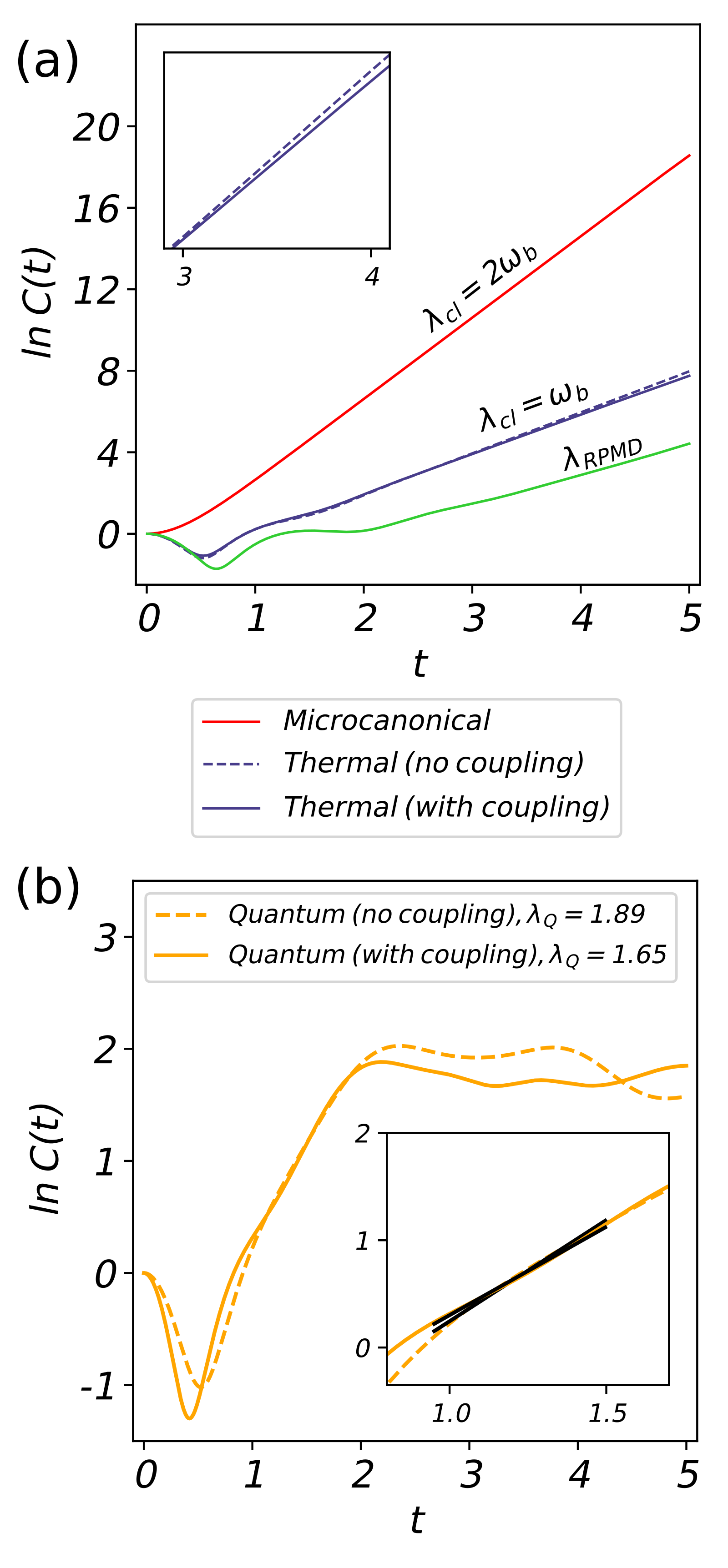}
    \caption{Illustration of quenching in the scrambling rates of the classical, semiclassical and quantum OTOCs. The microcanonical OTOC in (a) was obtained from a single classical trajectory initialised very close to the phase-space origin.
     The thermal OTOCs were computed at temperature $T=3T_c$, with $z$ of \eqref{Vcoupling} set to 0 (no coupling) and 2 (with coupling). }
    \label{fig:fig1}
\end{figure}
\fi

Fig.~\ref{fig1a} shows the computed classical OTOCs at $T=3T_c$ for $z=0$ (uncoupled) and $2$ (coupled). Both OTOCs behave similarly, with  exponential growth dominating after $t=2$, with a growth rate $\lambda_\text{cl}$ that is less than its long-time limit of $\lambda_\text{cl}=\omega_b=2$.  The greater reduction of lambda for $z=2$ ($\lambda_\text{cl}=1.75$) than $z=0$ ($\lambda_\text{cl}=1.92$)  indicates that the phase-space volume that dominates $C_\text{cl}(t)$ at these short times extends sufficiently far along the $y$-coordinate that switching on the coupling (which reduces the Hessian on moving away from the saddle along the $y$ coordinate) significantly reduces $\lambda_\text{cl}$. Fig.~\ref{fig1b} shows that there is a comparable reduction in $\lambda_Q$ for the quantum OTOC when $z$ is increased from $0$ to~$2$. These results suggest that, at least for this system at $T=3T_c$, a change in the coupling strength $z$ produces roughly equivalent changes in the distribution of Hessians that dominate the short time exponential growth of $C_\text{cl}(t)$ and  $C_\text{q}(t)$.

\if\figgy1
\begin{figure}[t]
\includegraphics[width=0.7\columnwidth]{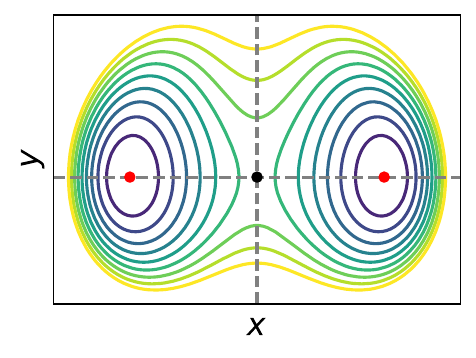}
    \caption{Schematic plot of the potential in \eqref{Vcoupling} for $z$=2, showing the minima (red dots) and the saddle (black dot) with the unstable mode along the $x$ direction. }
    \label{fig:fig2}
\end{figure}
\fi

An analogous mechanism also appears to explain why $\lambda_Q$ is significantly lower than $\lambda_\text{cl}$ even at $T>T_c$ (where there are no instantons and real-time coherence effects are relatively small \footnote{Real-time coherence effects become large at temperatures around $T_c$ and prevent exponential growth completely  below about $0.8T_c$.}).   In this case, the reduction in $\lambda_Q$ is caused by the spreading of the distribution along the quantum thermal fluctuations which are sampled by the quantum-Boltzmann distribution. We can represent this distibution in the usual imaginary-time (Euclidean-action) path integral representation, for which 
\begin{align}\label{cyclicPI}
    \text{Tr}[e^{-\beta \hat{H}}] = \oint \mathcal{D}\textbf{q}[\cdot] \: e^{-S_E[\textbf{q}(\cdot);\beta]\slash \hbar}
\end{align}
where 
\begin{align}
    S_E[\textbf{q}(\cdot);\beta] = \int_0^{\beta \hbar} d\tau \:\frac{m|\dot{\textbf{q}}(\tau)|^2}{2} + V(\textbf{q}(\tau))
\end{align}
is the Euclidean action and $\oint$ represents an integral over cyclic paths. A standard procedure \cite{feynman2018statistical} to quantify quantum fluctuations is to decompose the path space (and the Euclidean action) into centroid ($\textbf{Q}_0)$ and fluctuation components ($\boldsymbol{\xi}(\tau)$) to obtain
\begin{align}
    \text{Tr}[e^{-\beta \hat{H}}] = \int d\textbf{Q}_0 \oint_{\textbf{Q}_0} \mathcal{D}\boldsymbol{\xi}[\cdot] \: e^{-S_E[\boldsymbol{\xi}(\cdot);\beta]\slash \hbar}
\end{align}
It can be shown using perturbation theory that
\begin{align}\label{fluc_coupling}
    \frac{\partial^2 S_E}{\partial \textbf{Q}_0^2} = \beta\frac{\partial^2 V}{\partial \textbf{Q}_0^2}  + g\beta \: \langle |\boldsymbol{\xi}(\tau)|^2 \rangle_{\textbf{Q}_0} + ...
\end{align}
(the second term above is invariant with respect to $\tau$ by symmetry in imaginary time) which shows that the geometric centroid of the cyclic path is coupled quadratically (and symmetrically) to the fluctuation modes, analogously to the (classical) coupling between $x$ and $y$ in the model potential \eqref{Vcoupling}. This becomes more explicit if we decompose the cyclic paths as a fourier series $\textbf{q}(\tau) = \sum_{-\infty}^{\infty} \textbf{Q}_n e^{i\omega_n\tau}$, which yields $\langle |\boldsymbol{\xi}(\tau)|^2 \rangle_{\textbf{Q}_0} = \sum_{n\neq 0} |\textbf{Q}_n|^2$. 

\if\figgy1
\begin{figure}
\includegraphics[scale=0.8]{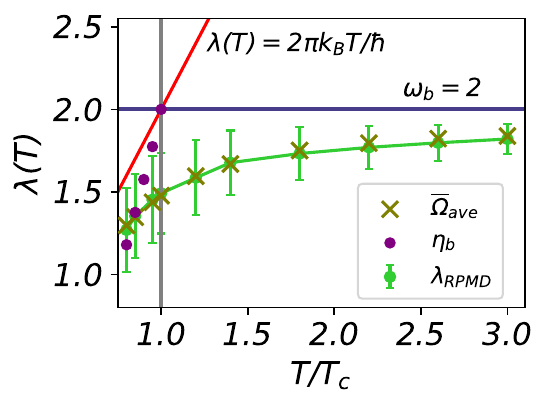}
    \caption{Comparison of $\lambda_\text{RPMD}$ with the short-time approximation $\overline\Omega_\text{ave}$ of \eqref{xiom} over a wide temperature range. Also plotted are the classical barrier frequency $\omega_b$, the instanton barrier frequencies $\eta_b$, and the MSS bound (red line).}
    \label{fig:fig3}
\end{figure}
\fi

To demonstrate the effect of fluctuation modes on the dynamics,
 we  follow ref.~\onlinecite{sadhasivam2023instantons} in computing the `ring-polymer molecular dynamics' (RPMD) OTOC
\begin{align}\label{rpmdOTOC}
    C_\text{RPMD}(t) = \frac{\hbar^2}{h^{2N}Z_N} \int d\textbf{q}^N \: d\textbf{p}^N \: e^{-\beta_N H_N} \left(\frac{\partial X_{0t}}{\partial X_0}\right)^2
\end{align}
in which
\begin{align}
X_0 = {1\over N} \sum_{k=1}^N x_k
\end{align}
is the $x$-centroid of $N$ replica phase-space points $({\bf p}^N;{\bf q}^N)\equiv \{p_1,\dots,p_N;q_1,\dots,q_N\}$ which are propagated classically using the hamiltonian
\begin{align}
    H_N &=  \frac{({\bf p}^N)^2}{2m} + U_N(\textbf{q}^N)\nonumber\\
    U_N(\textbf{q}^N) &= \sum_{i=1}^N V({\bf q}_i) + \frac{m} {2(\beta_N\hbar)^2}  \vert {\bf q}_i - {\bf q}_{i-1} \vert^2\label{RPHamil}
\end{align}
where $V$ is as defined in \eqref{Vcoupling} (with $z=2$), and  $U_N(\textbf{q}^N)\hbar$ is the discretised imaginary-time action obtained by trotterizing the quantum Boltzmann operator into $N$ imaginary time steps of length
$\beta_N=\beta/N$ (and $Z_N$ is the analogous partition function).
Clearly $C_\text{RPMD}(t)$ is an artificial classical construct which is not expected to reproduce the quantum OTOC $C(t)$
 (except in the limits $t\to0$ and $\beta\to 0$). However, $C_\text{RPMD}(t)$ gets one essential property right: the RPMD trajectories that dominate the OTOC at short times are drawn from the same distribution of Hessians around the saddle as those sampled by the exact quantum dynamics. In ref.~\onlinecite{sadhasivam2023instantons}, this property was shown to be sufficient to make $C_\text{RPMD}(t)$ obey the MSS bound, and to show how the quantum statistics impose the bound by shifting the saddle point on $U_N(\textbf{q})$ from the classical saddle geometry ${\bf q}^N=0$ at $T>T_c$ to the delocalised instanton geometry at $T<T_c$.

 Fig.~\ref{fig1a} plots $C_\text{RPMD}(t)$ at $3T_c$. There is no instanton at this temperature, but nonetheless the RPMD exponential growth rate $\lambda_\text{RPMD}$ (and $\lambda_q$, not shown) is significantly lower than $\lambda_\text{cl}$. This reduction can be explained analogously to the drop in $\lambda_\text{cl}$ on switching $z$ from 0 to 2, as arising because
 $ t$ is too short for the exponential growth of $C_\text{RPMD}(t)$ to have reached its $t\to\infty$ limit of $\omega_b= 2$. The exponential growth is dominated by a broad distribution of phase space around the saddle at ${\bf q}^N=0$, which, in this finite difference approximation, extends along the modes
 \begin{align}
{\bf Q}_n=(X_n,Y_n) = \sum_{k=1}^N  {\bf q}_k e^{i2n\pi k/N},\quad n=0,\dots, N-1
\end{align}
which  for $n\ne0$ describe quantum thermal fluctuations  around the centroid. Since the unstable mode at the saddle
lies along $X_0$ and does not rotate much on moving away from the saddle (but see below), the exponential growth of ${\partial X_{0t}}/{\partial X_0}$ around the saddle is determined mainly by (following \eqref{fluc_coupling})
\begin{align}
    -\frac{\partial^2 U_N({\bf q})}{\partial X_0^2} = m\omega_b^2 - \sum_{n=0}^{N-1}\left(  12g |X_n|^2 
    +z^2\alpha^2|Y_n|^2 \right)
\end{align}
The spread of the distribution along the ${\bf Q}_n$ modes therefore reduces $-{\partial^2 U_N({\bf q})}/{\partial X_0^2}$, which explains the reduction in $\lambda_\text{RPMD}$ with respect to $\lambda_\text{cl}$. As the temperature is decreased from $3T_c$ to $T_c$ (fig.~\ref{fig:fig3}), this reduction increases, since the amplitudes of the quantum fluctuations increases along ${\bf Q}_n$ as  $m\sum_n \omega_n^2 Q_n^2 /2$ (with $\omega_n = 2\sin(n\pi/N)/\beta_N\hbar$). Below $T_c$, the dependence of $-{\partial^2 U_N({\bf q})}/{\partial X_0^2}$ on ${\bf Q}_n$  becomes more complicated, but fig.~\ref{fig:fig3} shows that a comparable reduction of $\lambda_\text{RPMD}$ occurs with respect to its long time limit (i.e.~the instanton barrier frequency). 

\if\figgy1
\begin{figure}[t]
\labelphantom{fig4a}
\labelphantom{fig4b}
\includegraphics[scale=0.8]{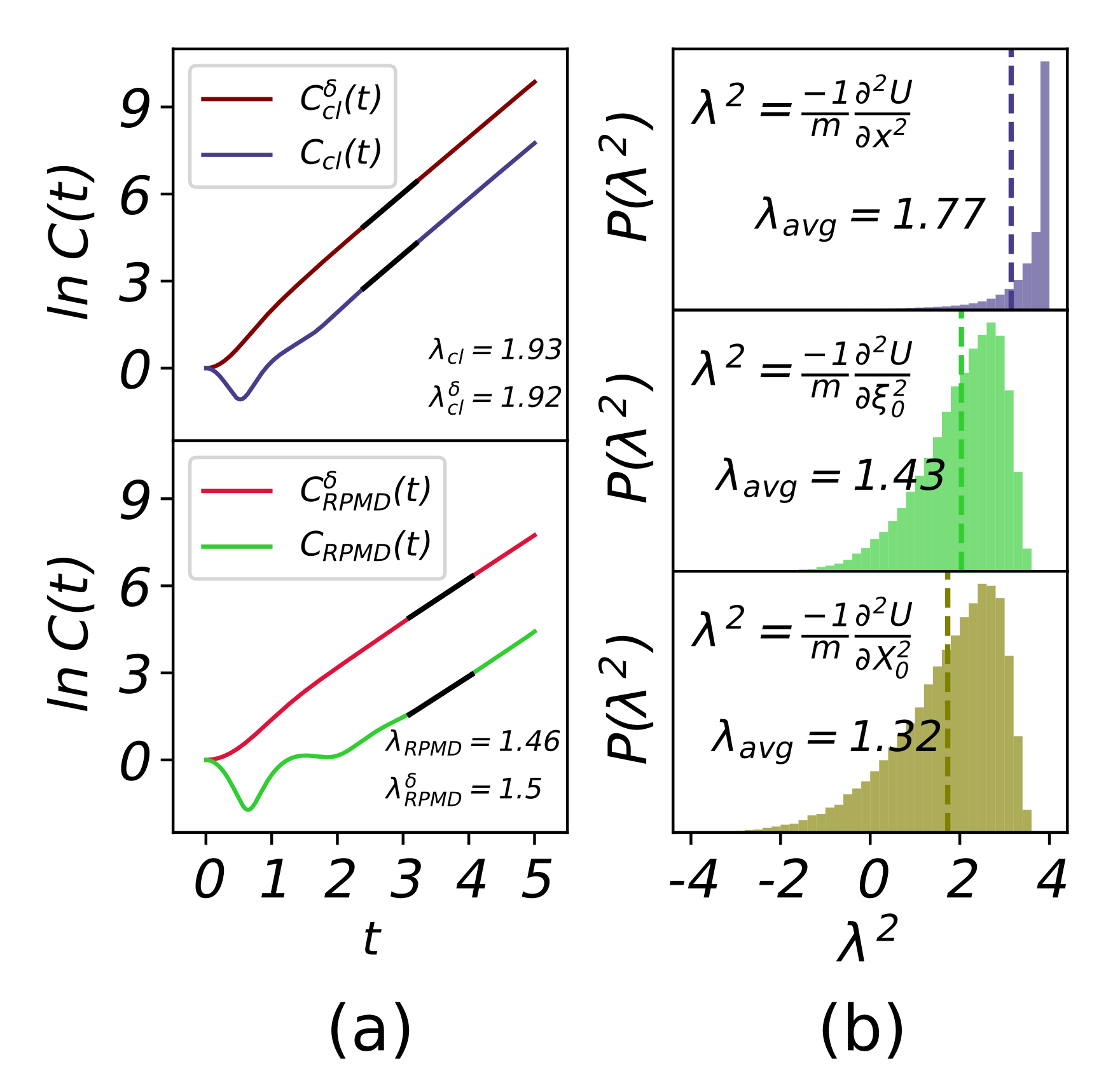}
    \caption{(a) Comparison of classical and RPMD OTOCs computed with an initial dividing-surface constraint ($\delta$-superscript) and without such a constraint (no $\delta$-superscript). The classical OTOCs were computed at $T=3T_c$; the RPMD at $T=0.95T_c$. (b) Distribution of the maximal (negative) Hessian eigenvalue $\lambda$ along the dividing surface in the classical (top) and RPMD simulations (middle), and of the projection of the latter along $X_0$ (bottom). The dashed vertical lines indicate the distribution averages $\lambda_\text{avg}$.}
    \label{fig:fig4}
\end{figure}
\fi

Having established that the exponential growth of $C_\text{cl}(t)$ and $C_\text{RPMD}(t)$ is slower at short times than in the  long time limit, we now ask whether the short-time growth can be estimated by taking a $t\to0$ limit. We have found that in general this is not true; but such a limit does seem to exist for potentials which, like $V({\bf q})$ of \eqref{Vcoupling}, have symmetric coupling between the stable and unstable modes of the saddle.  Fig.~\ref{fig4a} shows a plot of the quantity
\begin{align}\label{C0cl}
    C^\delta_\text{cl}(t) = \frac{1}{4\pi^2Z_{cl}} \int d\textbf{q} \: d\textbf{p} \: \delta(x) e^{-\beta H(\textbf{q},\textbf{p})} \left(\frac{\partial x_t}{\partial x}\right)^2_{\!(y,{\bf p})}
\end{align}
which is an OTOC in  which the initial distribution is constrained to a `dividing surface' along $y$ passing through $x=0$. By filtering out the large contributions from trajectories that originate in the wells, this OTOC grows exponentially from an earlier time than $C_\text{cl}(t)$, but at a very similar rate. In the limit $t\to 0$, $C^\delta_\text{cl}(t)$ grows quadratically, as $1+\omega_\text{ave}^2 t^2$, where  
\begin{align}
    \omega_\text{ave}^2 = -\frac{1}{h^2mZ_{cl}} \int d\textbf{q} \: d\textbf{p} \: \delta(x) e^{-\beta H(\textbf{q},\textbf{p})} {\partial^2 V({\bf q})\over \partial x^2}
\end{align}
is the average of the negative hessian over the dividing surface. Fig.~\ref{fig4a} shows that $\omega_\text{ave}$ is a much better approximation than $\omega_b$ to $\lambda_\text{cl}$. The accompanying histogram in fig.~\ref{fig4b} thus gives a good estimate of the distribution of negative hessian eigenvalues that dominate the exponential growth of $C_\text{cl}(t)$ at short times. 

Fig.~\ref{fig4a} also plots the analogous quantities 
\begin{align}\label{C0RP}
    C^\delta_{\text{RPMD}}(t) = \frac{\hbar^2}{h^{2N}Z_N} \int d\textbf{q}^N \: d\textbf{p}^N \:
    \delta({\bf Q}_0) e^{-\beta_N H_N} \left(\frac{\partial X_{0t}}{\partial X_0}\right)^2
\end{align}
and 
\begin{align}\label{lave_xi0_RPMD}
    \Omega_\text{ave}^2 = -\frac{1}{h^{2N}mZ_N} \int d\textbf{q}^N \: d\textbf{p}^N \:
    \delta({\bf Q}_0) e^{-\beta_N H_N} {\frac{\partial^2 U_N({\bf q}^N)}{\partial X_0^2}}
    \end{align}
 [where $\delta({\bf Q}_0):=\delta(X_0)\delta(Y_0)$]
for the RPMD OTOC. As for the classical OTOC, the exponential growth of $C^\delta_N(t)$ is a close approximation to that of $C_N(t)$; the hessian average $\omega_\text{ave}^2$ is less good, but this is because we need to account for the small rotations  $X_0\to \xi_0$ of the hessian unstable eigenvector on moving away from the saddle; fig.~\ref{fig:fig3} shows that  
\begin{align}\label{xiom}
    \overline\Omega_\text{ave}^2 = -\frac{1}{h^{2N}mZ_N} \int d\textbf{q}^N \: d\textbf{p}^N \:
    \delta({\bf Q}_0) e^{-\beta_N H_N} {\frac{\partial^2 U_N({\bf q}^N)}{\partial \xi_0^2}}
    \end{align}
gives an excellent approximation to $\lambda_\text{RPMD}$ across the full temperature range tested (including below below $T_c$). Note that the maximum negative hessian eigenvalue in the quantum Boltzmann distribution about $(X_0,Y_0)=(0,0)$ (see fig.~\ref{fig4b}) is peaked below $\omega_b$ (because pairs of fluctuation modes $X_{\pm n},Y_{\pm n}$ are doubly degenerate); this explains why quantum fluctuations give such a significant reduction in the short-time exponential growth-rate of $C_\text{RPMD}(t)$.
 
We emphasise that the short-time approximations $\Omega_\text{ave}$ and $\overline\Omega_\text{ave}$ appear to work only in the special case that the intermode coupling in $V({\bf q})$ is symmetric (about the coordinates $x$ and $y$ in this case). For example, if the $xy$-coupling term in \eqref{Vcoupling} is replaced by the asymmetric coupling used in ref.~\onlinecite{sadhasivam2023instantons}, the resulting $\overline\Omega_\text{ave}$ is not a good approximation to $\lambda_\text{RPMD}$.  

Although this article is concerned with the scrambling rate of OTOCs, we note that the findings above may help to understand some of the properties of the multi-time correlation functions used in non-linear spectroscopy, which usually contain commutators such as $[\hat q_t,\hat p]$ (or the corresponding $i\hbar\,\partial q_t/\partial q$ in the classical limit) \cite{dellago2003simulation,mukamel1996classical,hamm2019velocity}. An interesting question is why none of these functions appear to grow exponentially, not even in the classical limit. For the simplest of these examples, which contain a single power of $i\hbar\,\partial q_t/\partial q$, integration by parts shows that any exponential growth must cancel out. However, some of these functions contain multiple powers of $i\hbar\,\partial q_t/\partial q$ (evaluated at different times) for which the lack of exponential growth cannot be so easily explained away \cite{hamm2019velocity}. It may be that quenching processes similar to those discussed above prevent exponential growth.

In conclusion, the scrambling rate over an isolated saddle is reduced by a hierarchy of processes. First, escape from the  neighbourhood of the saddle reduces the rate by a factor of two. Second, the quantum OTOC grows exponentially only at short times, which means that the range of negative Hessian eigenvalues contributing to the scrambling rate is roughly as wide as the thermal distribution around the saddle \footnote{The short-time reduction appears to   be closely related to the multifractal distribution of Lyapunov exponents reported in ref.~\onlinecite{pappalardi2023quantum}}. This  broad distribution slows down the overall growth rate of the classical OTOC and still more so of the quantum OTOC on account of quantum thermal fluctuations.  At temperatures below $T_c$, the quantum scrambling rate is further reduced by instanton formation, in line with the MSS bound. Finally, the quantum scrambling rate is also affected by real-time coherence which is difficult to predict but typically reduces it.  The short-time reduction in the scrambling rate makes it very unlikely that any system with an isolated saddle point can saturate the MSS bound.

The authors acknowledge funding from the Leverhulme Trust (VGS and SCA), the UK Engineering and Physical Sciences Research Council (ACH) and the Deutsche Forschungsgemeinschaft (YL). VGS also acknowledges support from St John's College, Cambridge.

\bibliography{Citations}
\end{document}